# A Step towards Software Corrective Maintenance: Using RCM model

Shahid Hussain
Department of computing
Namal College
Mianwali, Pakistan
Shahidhussain2003@yahoo.com

Muhammad Zubair Asghar
Department of ICIT
Gomal University
Dera Ismail Khan, Pakistan
zubair_icit@yahoo.com

Bashir Ahmad
Department of ICIT
Gomal University
Dera Ismail Khan, Pakistan
bashahmad@gmail.com

Shakeel Ahmad
Department of ICIT
Gomal University
Dera Ismail Khan, Pakistan
Shakeel_1965@yahoo.com

*Abstract*--From the preliminary stage of software engineering, selection of appropriate enforcement of standards remained a challenge for stakeholders during entire cycle of software development, but it can lead to reduce the efforts desired for software maintenance phase. Corrective maintenance is the reactive modification of a software product performed after delivery to correct discovered faults. Studies conducted by different researchers reveal that approximately 50 to 75% of the effort is spent on maintenance, out of which about 17 to 21% is exercised on corrective maintenance. In this paper, authors proposed a RCM (Reduce Corrective Maintenance) model which represents the implementation process of number of checklists to guide the stakeholders of all phases of software development. These check lists will be filled by corresponding stake holder of all phases before its start. More precise usage of the check list in relevant phase ensures successful enforcement of analysis, design, coding and testing standards for reducing errors in operation stage. Moreover authors represent the step by step integration of checklists in software development life cycle through RCM model.

## I. INTRODUCTION

The selection of proper enforcement of standards is the challenging task right from early stage of software engineering which has not got definite importance by the concerned stakeholders. Software maintenance takes more effort than all other phases of software life cycle, but it has not been given as much importance as it deserved. It is an admitted fact that approximately 60 to 70% effort is spent on maintenance phase of software development life cycle. Software maintenance is classified into corrective, adaptive, perfective and preventive maintenance. According to IEEE[2, 3], corrective maintenance is the reactive modification of software product performed after delivery to correct discovered faults, adaptive maintenance is the modification of a software product performed after delivery to keep software usable in a changed or changing environment, perfective maintenance is the modification of a software product after delivery to improve performance or maintainability and preventive maintenance is performed for the purpose of preventing problems before they occur. In this paper the main focus of authors is towards corrective maintenance to overcome the all problems arising in requirements, design, coding, documentation and testing activities.

According to Yogesh [1] software maintenance process is costs 50% for Perfective maintenance, 25% for Adaptive maintenance, 21% for Corrective maintenance and 4% for Preventive maintenance. In this paper authors proposed a RCM model to reduce the maintenance cost by incorporating checklists for concerned stakeholder of each phase of software development life cycle. This would lead to reduction of post efforts made by stake holders during corrective maintenance and decrease the percentage effort of corrective maintenance suggested by Yogesh[ 1 ].

## II. SOFTWARE MAINTENANCE

Software maintenance is the process to correct the faults arises in software product after its delivery. IEEE [2, 3] definition for software maintenance is:

*The modification of a software product after delivery to correct faults, to improve performance or other attributes or to adapt the product to a modified environment.*

It has been observed during different studies that software maintenance is the most time consuming activity in SDLC, Fig-1 shows maintenance iceberg depicting the time consuming nature of software maintenance. Software is to be modified when it is not fulfilling the needs of the environment in which it works.

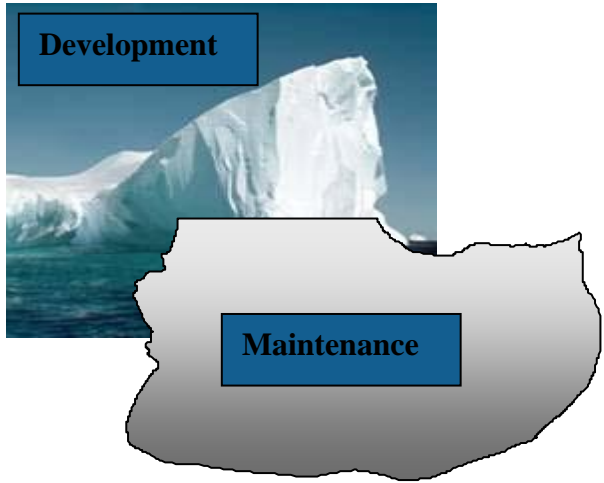

Figure-1. Maintenance Ice Berg, [Martin and McClure 1983]

Different models and techniques are proposed by researchers in the area of software corrective maintenance [4, 5, 6]. Walia and Jeffrey proposed a catalog C[7] for aid of developers to reduce errors during the requirement inspection process and to improve overall software quality. The Study of Jie-Cherng Chen and Sun-Jen Huang [8] show the empirical evidence for all those problem factors which are faced during the software development phase and affects the software maintainability negatively. Similarly, Andrea De Lucia et ,al [9] provided an empirical assessment and improvement of the effort estimation model for corrective maintenance. Authors' proposed model provides an easy and sequential procedure for integrating checklists into SDLC for reducing effort for software corrective maintenances

### III. RCM Model

The whole work of software development life cycle is dived into five main phases such as requirement elicitation, requirement specification, designing, coding and testing. In each phase if roles are not properly guided to operate their activities then it can cause to increase the efforts required for maintenance of software especially for corrective maintenance. In this paper authors use RCM model to provide guidelines to concerned stakeholders of each phase of software development life cycle. The working of RCM model is represented through Figure-1. Before the start of each phase concerned stack holders fill a checklist which guides them about standard methods to perform their activities. If all concerned stakeholders of each phase worked according to the guidelines of checklist then it can affect the total effort required for software corrective maintenance. The stakeholders of requirement elicitation phase will fill the checklist shown in Table-1 before start of their work. The evaluation result of this checklist will show that all requirements are clear and understandable to concerned stakeholders. This would lead to reduce the error chances which can arise due to ambiguities in requirement elicitation process. The stakeholders of requirement specification phase will fill the checklist shown in Table-2 before the start of their work. The evaluation result of this checklist will show that specification of requirements is understandable to the concerned stakeholders and reduces the error chances which can arise due to improper specification of requirements. The stakeholders of designing phase will fill the checklist shown in Table-3 before the start of their work. The evaluation result of this checklist will show that the architectural, data, procedural and user interface designing of software is understandable to the concerned stack holders and reduces the error chances which can arise due to lack of proper understanding of designing activities. The stakeholders of coding phase will fill the checklist shown in Table-4 before the start of their work. The evaluation result of this checklist will show that coding standard features are understandable to concerned stakeholders and reduces the error chances which can arise due to lack of proper understanding of coding constructs. The stakeholders of testing phase will fill the checklist shown in Table-5 before the start of their work. The evaluation result of this checklist will show that software will be tested with respect to each aspect and reduces the error chances which can arise due to improper testing process.

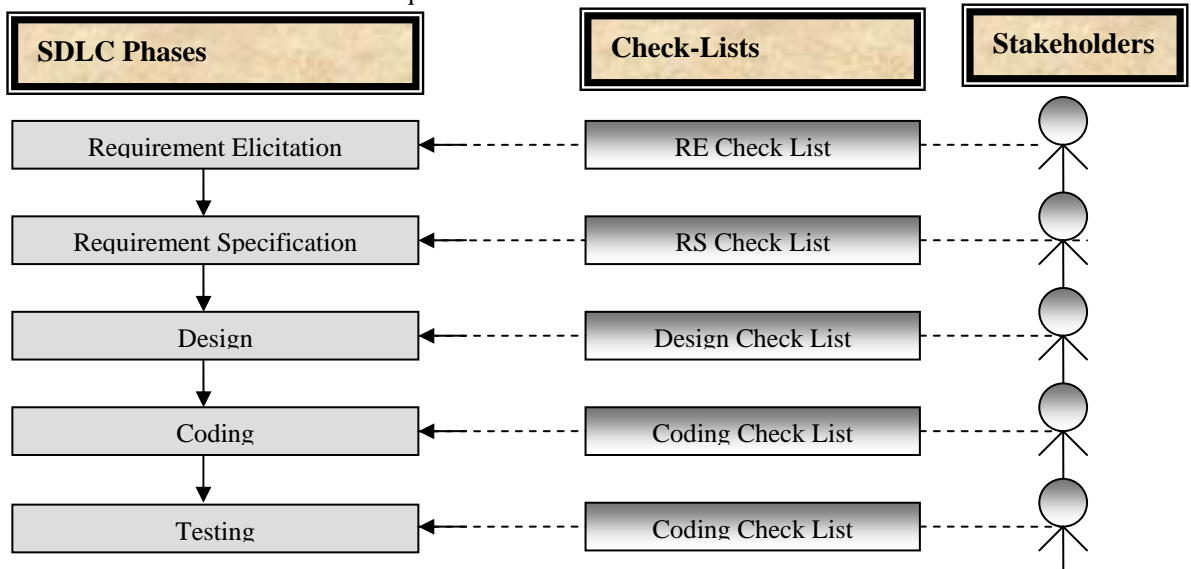

Figure- 2. RCM Model for Software Corrective Maintenance

The checklist for requirement elicitation phase enables the concern stakeholders to identify the requirements in a clear and precise way. Repetition in gathered requirement should be avoided, easy to understand in recommended natural. Moreover dependencies among requirements should be clear and once requirement elicitation is completed, then further no requirement can be gathered. If concern stakeholder such as system analyst follows this checklist in precise manner then errors which can arise due to inconsistencies and repetition can be avoided and it will directly impact the corrective maintenance efforts. The column heading *Yes* and *No* of Table-1 show that given points of checklist are clearly understandable by concern stakeholders or not. And the checklist will be analyzed on the base of these values. Moreover, same concept is used for other checklists.

TABLE-1. CHECKLIST FOR STAKEHOLDERS OF REQUIREMENT ELICITATION PHASE

| ACTIVITY CODE | DESCRIPTION | YES | NO |
|---|---|---|---|
| RE-1 | Natural Language for requirement gathering is understandable. | | |
| RE-2 | No requirement would be repeated | | |
| RE-3 | Each requirement should be clear and accurate | | |
| RE-4 | The source of each dependent requirement should identifiable. | | |
| RE-5 | All sources to collect requirement should be known able. | | |
| RE-6 | Take full detail of each requirement from customer | | |
| RE-7 | No requirement of customer will entertain after collecting all requirements and starting of new phase | | |

The checklist for requirement specification phase enables the concern stakeholders to use proper methods for specification of requirements. It ensures that SRS should be clear and understandable to all stakeholders. The stakeholders of this phase should have sufficient knowledge of formal and informal specification and its tools or languages.

TABLE-2. CHECKLIST FOR STAKEHOLDERS OF REQUIREMENT SPECIFICATION PHASE

| ACTIVITY CODE | DESCRIPTION | YES | NO |
|---|---|---|---|
| RS-1 | The structure of SRS is clear and understandable | | |
| RS-2 | Knowledge of informal specification of requirements | | |
| RS-3 | Knowledge of formal specification of requirements | | |
| RS-4 | Use of informal specification tool or language | | |
| RS-5 | Use of formal specification tool or language | | |
| RS-6 | SRS must be clear to all stack holders | | |
| RS-7 | Data, functional and behavioral modeling is understandable | | |

The checklist for designing phase enables the concern stakeholders to perform both back-end and front-end designing of softwares in precise form. This checklist leads to make easy and understandable transformation process of analysis model into different types of designing models such as architectural, data, procedural and user interface designing. The relationship among modules should be clear and understandable for all stakeholders.

TABLE-3. CHECKLIST FOR STAKEHOLDERS OF DESIGN PHASE

| ACTIVITY CODE | DESCRIPTION | YES | NO |
|---|---|---|---|
| D-1 | SRS is clear and understandable | | |
| D-2 | Architectural design of software is clear and users have performed acceptance testing | | |
| D-3 | Black box testing on architectural design have been performed | | |
| D-4 | Database designing is properly designed and understandable | | |
| D-5 | Relationship among dependent modules is clear | | |
| D-6 | User interface designing is accepted by user | | |
| D-7 | Data Dictionary is clear and properly designed | | |
| D-8 | Design strategy either top-down or bottom-up is clear | | |
| D-9 | Standards for procedural designing are clear | | |

The checklist for coding phase enables the concern stakeholders to clearly understand the basic construct such variable, array and functions of programming language. Moreover, this checklist shows that validation process of text and exception handling process will be clear to concern stakeholders.

TABLE-4. CHECKLIST FOR STAKEHOLDERS OF CODING PHASE

| ACTIVITY CODE | DESCRIPTION | YES | NO |
|---|---|---|---|
| C-1 | Each variable should be correctly typed. | | |
| C-2 | Used data structure should be clear | | |
| C-3 | Scope of all variables should be clear | | |
| C-4 | Variables are initialized without garbage values | | |
| C-5 | Size of buffers is appropriated. | | |
| C-6 | Buffer's overflow are properly checked | | |
| C-7 | Signatures of function are understandable | | |
| C-8 | Functions should be properly called | | |
| C-9 | Use of formal and actual parameters should be clear | | |
| C-10 | Recursive function should be properly called and ended | | |
| C-11 | All other construct of programming language should be properly used | | |
| C-12 | Use of third party control is valid. | | |
| C-13 | All database files should be proper open or close when control is transfer from one module to another. | | |
| C-14 | Proper validation rules and validation text should be defined | | |
| C-15 | Exception handling should be properly embedded into program structure | | |

The checklist for testing phase enables the concern stakeholders to clearly understand the testing methods such as white-box, grey-box and black-box. Moreover, this checklist presents that all testing activities will be done properly and understandable to all stakeholders.

TABLE-5. CHECKLIST FOR STAKEHOLDERS OF TESTING PHASE

| ACTIVITY CODE | DESCRIPTION | YES | NO |
|---|---|---|---|
| T-1 | Unit testing for each component should be properly performed | | |
| T-2 | Module level testing should be properly performed | | |
| T-3 | Modules are properly integrated and tested | | |
| T-4 | Function of each module should be tested through functional testing | | |
| T-5 | In white-box testing, each path should be clearly defined | | |
| T-6 | Use of all constructs of programming language should be properly tested. | | |
| T-7 | Functional requirement of users should be tested | | |

## IV IMPLEMENTATION PLAN

The implementation process of RCM model has been started shown in fig-3. Two teams of students are used to develop same project. The development experience level of all students of both teams is same. The development and maintenance process for first project will be ordinary but in second project development and maintenance team will follow the rules of RCM model and will analyzed the result. The stakeholders of team, which are using RCM model, are trained to understand the purpose of checklist. For example if a programmer can not understand the function of use of buffer, multi threading, recursive calling, parameters' scope and access, or multi tasking then he cant not fill the related checklist effectively. Before start of project, only the stakeholders of requirement elicitation phase will be trained. When Requirement Elicitation will going to end then parallel training of next phase stakeholders will be started and this process will remains continue till the end of software' development. This strategy will helps to reduce the extra time consumed on stakeholders' training. The project manager will be responsible to overlook the work of both projects and analyze the result.

Different factors are targeted to analyze the performance of RCM model such as quality, defect rates, reduction in efforts, cost, complexity, productivity and reliability.

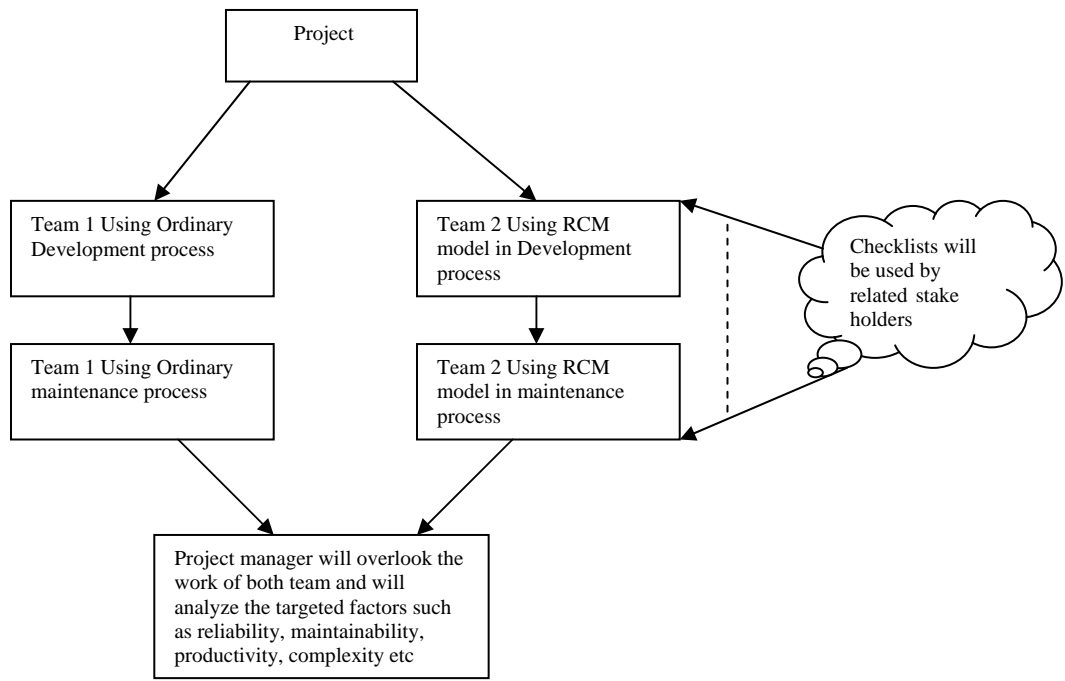

Figure-3. Flow graph for Implementation Strategy

## V. CONCLUSION

Software maintenance process consumes half of the budget and time to complete a software project and usually 21% of total maintenance efforts are devour by corrective maintenance. The corrective maintenance efforts are increases due to flaws remains in other phase of software development life cycle. These flaws can be overcome if stakeholders fully understand the activities of each concern phase. Authors proposed a RCM model which comprises on filling and analyzing process of checklists in each phase. If all stakeholders of each phase filled the checklist in precise manner then evaluated result of each checklist shows that how much stakeholder have understand the activities. Such process would leads to reduce the corrective maintenance effort which is increasing the overall effort percentage of software maintenance. RCM model is in its infancy period, it just presents an idea of how to reduce software corrective maintenance effort. Moreover, the checklist of RCM can be updated by stakeholder who will apply this model during development process of software.

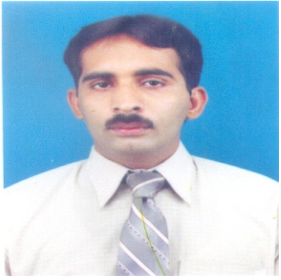

**Mr. Shahid Hussain** has done MS in Software Engineering from City University, Peshawar, Pakistan. He has got distinction throughout his academic carrier. He has done his research by introducting best practices in different software process models. I have introduced a new role communication model in RUP using pairing programming as best practise.. Recently, I am working as coursechair cum Leccturer in Namal College, an associate college of University of Bradford. Moreover, I have publish many research paper in different national/international journals and conferences such as MySec04, JDCTA, IJCSIS, NCICT, ZABIST.

Similarly, I have worked as computer programmer with SRDC (British Council), Peshawar, Pakistan and have developed many softwares. My furtue aim is to join an organization where I can polish my abilities.

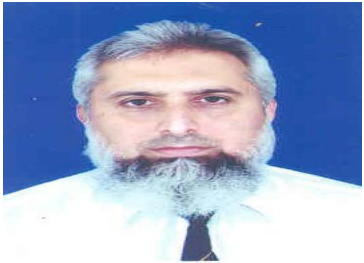

**Dr. Shakeel Ahmad** received his B.Sc. with distinction from Gomal University, Pakistan (1986) and M.Sc. (Computer Science) from Qauid-e-Azam University, Pakistan (1990). He served for 10 years as a lecturer in Institute of Computing and Information Technology (ICIT), Gomal University Pakistan. Now he is serving as an Assistant Professor in ICIT, Gomal University Pakistan since 2001. He is among a senior faculty member of ICIT. Mr. Shakeel Ahmad received his PhD degree (2007) in Performance Analysis of Finite Capacity Queue under Complex Buffer Management Scheme.

Mr. Shakeel's research has mainly focused on developing cost effective analytical models for measuring the performance of complex queueing networks with finite capacities. His research interest includes Performance modelling, Optimization of congestion control techniques, Software refactoring, Network security, Routing protocols and Electronic learning. He has produced many publications in Journal of international repute and also presented papers in International conferences.

**Mr. Muhammad Zubair Asghar** is an MS student in Institute of Computing and information technology, Gomal University D.I.Khan, Pakistan. He has got distinction throughout his academic carrier. He is doing specialization in the area of software corrective maintenance. Author has also done work in the area of Artificial intelligence and got published two international publications in the areas of Robot simulation and medical expert systems.